\documentclass[12pt]{article}
\usepackage{epsf,latexsym}
\usepackage{amsfonts,amssymb} 
\epsfverbosetrue
\usepackage[ps,arc,frame]{xy}

\newcommand{\rxyg}[2]{{\begin{xy} 0;<2mm,0mm>:<0mm,2mm>::0;0,
,(5,-2)*{a} ,(10,-1.8)*{b} ,(15,-2)*{c} ,(20,-1.8)*{d} ,(2,-5)*{a}
,(1.8,-10)*{b} ,(2,-15)*{c} ,(1.8,-20)*{d} ,(5,-5)*\cir(#1,0){}
,(10,-5)*\cir(#1,0){} ,(15,-5)*\cir(#1,0){} ,(20,-5)*\cir(#1,0){}
,(5,-10)*\cir(#1,0){} ,(10,-10)*\cir(#1,0){} ,(15,-10)*\cir(#1,0){}
,(20,-10)*\cir(#1,0){} ,(5,-15)*\cir(#1,0){} ,(10,-15)*\cir(#1,0){}
,(15,-15)*\cir(#1,0){} ,(20,-15)*\cir(#1,0){} ,(5,-20)*\cir(#1,0){}
,(10,-20)*\cir(#1,0){} ,(15,-20)*\cir(#1,0){} ,(20,-20)*\cir(#1,0){}
#2\end{xy}}}

\newcommand{\double}[1]{\mathbb{#1}}

\newcommand{\cc}{\double{C}}

\newcommand{\rr}{\double{R}}
\newcommand{\zz}{\double{Z}}
\newcommand{\qqq}{\double{Q}}

\newcommand{\aaa}{\mathcal{A}}

\newcommand{\hhh}{\double{H}}
\newcommand{\mm}{\mathcal{M}}

\newcommand{\pp}{\pmatrix}

\newcommand{\ul}[1]{\underline{#1}}
\newcommand{\ot}{\otimes}
\newcommand{\op}{\oplus}

\newcommand{\bb}{\begin{eqnarray}}
\newcommand{\ee}{\end{eqnarray}}
\newcommand{\eee}{\nonumber\end{eqnarray}}
\newcommand{\qq}{\quad}
 
\begin{document}

\font\twelve=cmbx10 at 13pt
\font\eightrm=cmr8

\thispagestyle{empty}

\begin{center}

CENTRE DE PHYSIQUE TH\'EORIQUE \footnote{\,  Unit\'e Mixte de Recherche  (UMR 6207)
 du CNRS  et des Universit\'es Aix--Marseille 1 et 2\\
${}$\qq\qq\qq et  Sud
Toulon--Var, Laboratoire affili\'e \`a la FRUMAM (FR 2291)} 
\\ CNRS--Luminy, Case 907\\ 13288 Marseille Cedex 9\\
FRANCE\\ 

\vspace{2cm}

{\Large\textbf{On a Classification of Irreducible \\
 Almost Commutative
Geometries III}} \\

\vspace{1.5cm}

{\large Jan--Hendrik Jureit
\footnote{\, and Universit\'e de Provence and Universit\"at Kiel,\\
${}$\qq\qq\qq\ jureit@cpt.univ-mrs.fr}, 
Thomas Sch\"ucker
\footnote{\, and Universit\'e de Provence,\\
${}$ \qq\qq
\qq schucker@cpt.univ-mrs.fr },
Christoph Stephan
\footnote{\, and Universit\'e de Provence and Universit\"at Kiel,\\
${}$\qq\qq\qq\ stephan@cpt.univ-mrs.fr} }

\vspace{1.5cm}

{\large\textbf{Abstract}}
\end{center}
We extend a classification of irreducible, almost commutative
geometries whose spectral action is dynamically non-degenerate to
internal algebras that have four simple summands.

\vspace{1.5cm}

\vskip 1truecm

PACS-92: 11.15 Gauge field theories\\
\indent MSC-91: 81T13 Yang-Mills and other gauge theories

\vskip 1truecm

\noindent CPT-2005/P.015\\

\vspace{2cm}

\section{Introduction}

A Yang-Mills-Higgs model is specified by choosing a real compact
Lie group describing the gauge bosons and three unitary
representations describing the left- and right-handed fermions and
the Higgs scalars. Connes \cite{con} remarks that the set of all
Yang-Mills-Higgs models comes in two classes, fig.1. The first is tiny
and contains all those models that derive from gravity by generalizing
Riemannian to almost commutative geometry. The intersections of
this tiny class with the classes of left-right symmetric, grand unified
or supersymmetric Yang-Mills-Higgs models are all empty. However
the tiny class does contain the standard model of electromagnetic,
weak and strong forces with an arbitrary number of colours. 

\begin{figure}[h]
\label{versus}
\epsfxsize=11cm
\hspace{2.1cm}
\epsfbox{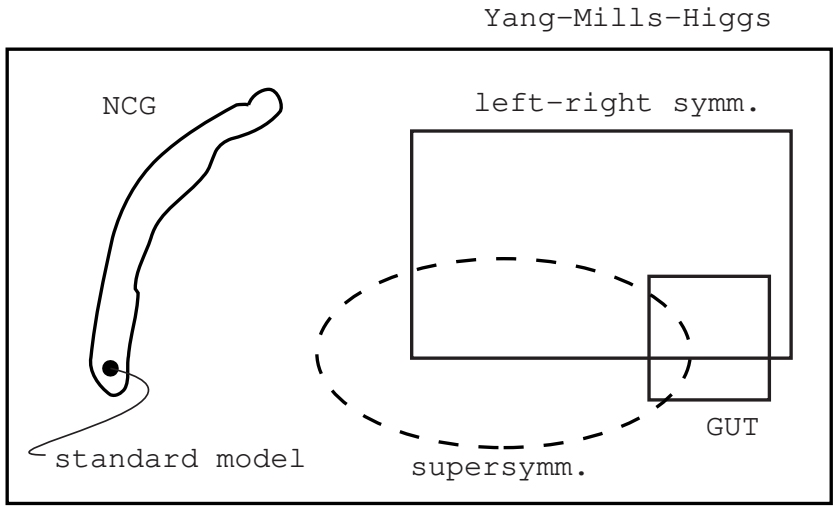}
\caption{Pseudo forces from noncommutative geometry}
\end{figure}

The first class is tiny, but still
infinite and difficult to assess. We have started putting some order
into this class. The criteria we apply are heteroclitic \cite{class}. 

Two criteria are simply motivated by simplicity: we take the internal
algebra to be simple, or with two, three,... simple summands, we
take the internal triple to be irreducible. 

Two criteria are motivated
from perturbative quantum field theory of the non-gravita\-tional
part in flat timespace: vanishing Yang-Mills anomalies and
dynamical non-degen\-er\-acy. The later imposes that the number
of possible fermion mass equalities be restricted to the minimum and
that they be stable under renormalization flow. The origin of these
mass equalities is the following. In almost commutative geometry, the
fermionic mass matrix is the `Riemannian metric' of internal space
and as such becomes a dynamical variable. Its `Einstein equation' is the
requirement that the fermionic mass matrix minimize the Higgs
potential. Indeed, in almost commutative geometry, the Higgs potential
is the `Einstein-Hilbert action' of internal space. 

Two criteria are motivated from particle phenomenology: We want
the fermion representation to be complex under the little group in
each of its irreducible components, because we want to 
distinguish particles from anti-particles by means of unbroken
charges. We want possible massless fermions to remain neutral under
the little group, i.e. not to couple to massless gauge bosons.

Two criteria are motivated from the hope that, one day, we will have
a unified quantum theory of all forces: vanishing mixed gravitational
Yang-Mills anomalies and again dynamical non-degeneracy. 

Our analysis is based on a classification of all finite spectral triples
by means of Krajewski diagrams \cite{kraj} and centrally extended
spin lifts of the automorphism group of their algebras \cite{fare}.
The central extension serves two purposes: (i) It makes the spin lift of
complex algebras $M_n(\cc)$ well defined. (ii) It
allows the corresponding $U(n)$ anomalies to cancel. 

Of
course we started with the case of a simple internal algebra. Here
there is only one contracted, irreducible Krajewski diagram and all
induced Yang-Mills-Higgs models are dynamically degenerate. 

The second case concerns algebras with two simple
summands. It admits three contracted, irreducible Krajewski diagrams,
one of them being a direct sum of two diagrams from the first case.
The corresponding triples induce two Yang-Mills-Higgs models which
are dynamically non-degenerate and anomaly free:
$SU(2)$ with a doublet of left-handed fermions and a
singlet of right-handed fermions. The Higgs scalar is a doublet.
Therefore the little group is trivial, $SU(2)\rightarrow \{1_2\}$,
and the fermion representation is real under the little group. The
second model is the $SO(2)$ submodel of the first. 

The case of three summands has 41 contracted, irreducible 
diagrams plus direct sums. Its combinatorics is on the limit of what
we can handle without a computer. There are only four induced
models satisfying all criteria: The first is the standard model with $C$
colours, $C\ge 2$,
$SU(2)\times U(1)\times SU(C)$, with one generations of leptons and
quarks and one colourless $SU(2)$ doublet of scalars. The three others
are submodels,
$SO(2)\times U(1)\times SU(C)$,
$SU(2)\times U(1)\times SO(C)$ and $SU(2)\times U(1)\times
USp(C/2)$ and must have $C\ge 2$, the last submodel of course
requires $C$ even, $C\ge 4$. 

In the following we push our classification to four summands. For
simplicity we will only consider diagrams made of letter changing
arrows, i.e. we exclude arrows of type $a\ \bar a$ because in the first
three cases these arrows always produced degeneracy. When the
algebra is simple, of course all arrows are of this type and all models
were degenerate. All direct sum diagrams in cases two and three
necessarily contain such arrows.  For two summands, there is only
one contracted, irreducible diagram, that is not a direct sum and that
is made of letter changing arrows,  for three summands there are 30
such diagrams.

For four summands, a well educated computer \cite{js} tells us that there
are 22 contracted, irreducible diagrams made of letter changing arrows
and which are not  direct sums. They are shown in fig.2. We have two
pleasant surprises: (i) the number of these diagrams is small, (ii) there
are only two ladder diagrams, diagrams 18 and 19, i.e. diagrams with
vertically aligned arrows, and all other diagrams have no subdiagrams of
ladder form. We will see that these other diagrams are easily dealt with.

\section{Statement of the result}

Consider a finite, real, $S^0$-real, irreducible
spectral triple whose algebra has four simple summands
and the extended lift as described in \cite{fare}. Consider the list of all
Yang-Mills-Higgs models induced by these triples and lifts. Discard
all models that have either 
\begin{itemize}\item
 a dynamically
degenerate fermionic mass spectrum,
\item
 Yang-Mills or gravitational
anomalies, 
\item
 a fermion multiplet whose representation under
the little group is real or pseudo-real, 
\item
or a massless fermion transforming
non-trivially under the little group.
\end{itemize}
 The remaining models are the
following, $C$ is the number of colours, the gauge group is on the
left-hand side of the arrow, the little group on the right-hand side:
\begin{description}
\item[$C=3,5...$]
\bb\frac{SU(2)\times U(1)\times SU(C)}{\zz_2\times \zz_C}&
\longrightarrow 
&\frac{U(1)\times SU(C)}{\zz_C}\eee
The left-handed fermions transform according to a multiplet 
${\ul 2}\ot {\ul C}$ with hypercharge $q/(2C)$ and a multiplet 
$ {\ul C}$ with hypercharge $-q/2$. The right-handed
fermions sit in two multiplets ${\ul C}$ with hypercharges
$q(1+C)/(2C)$ and $q(1-C)/(2C)$ and one singlet with hypercharge
$-q$, $q\in\qqq$. The elements in $\zz_2\times \zz_C$ are
embedded in the center of $ SU(2)\times U(1)\times SU(C)$ as
\bb \left( \exp{\frac{2\pi i k}{ 2  }}
\,1_2\,,\, \exp[2\pi i(Ck-2\ell)/q]\,,\, \exp{\frac{2\pi i \ell}{ C 
}}\,1_p\right),\qq  k=0,1,\qq  \ell =0,1,...,C-1.\eee
The Higgs scalar transforms as an
$SU(2)$ doublet,
$SU(C)$ singlet and has hypercharge $-q/2$.

With the number of colours $C=3$, this is the standard model with one
generation of quarks and leptons. 

We also have in our list two submodels of the above model defined
by the subgroups
\bb\frac{SO(2)\times U(1)\times SU(C)}{\zz_2\times \zz_C}&
\longrightarrow 
&\frac{U(1)\times SU(C)}{\zz_C},\cr \cr \cr 
\frac{SU(2)\times U(1)\times SO(C)}{\zz_2}&
\longrightarrow 
&U(1)\times SO(C).\eee
They have the same particle content as the standard model, in the
first case only the $W^\pm$ bosons are missing, in the second case
roughly half the gluons are lost.
\item[$C=2,4...$]
\bb\frac{SU(2)\times U(1)\times SU(C)}{\zz_C}&
\longrightarrow 
&\frac{U(1)\times SU(C)}{\zz_C}\eee
with the same particle content as for odd $C$.  But now we have three
possible submodels:
\bb\frac{SO(2)\times U(1)\times SU(C)}{\zz_C}&
\longrightarrow 
&\frac{U(1)\times SU(C)}{\zz_C}\,,\cr \cr \cr 
{SU(2)\times U(1)\times SO(C)}&
\longrightarrow 
&{U(1)\times SO(C)},\cr \cr \cr 
\frac{SU(2)\times U(1)\times USp(C/2)}{\zz_2}&
\longrightarrow 
&\frac{U(1)\times USp(C/2)}{\zz_2}.\eee
\item[electro-strong]
\bb{ U(1)\times SU(C)}&
\longrightarrow 
&{U(1)\times SU(C)}\eee
The fermionic content is ${\ul C}\op {\ul 1}$, one quark and
one charged lepton.  The two
fermion masses are arbitrary but different and nonvanishing.
The number of colours is greater than or equal to two, the two charges
are arbitrary but vectorlike, of course the lepton charge must not
vanish. There is no scalar and no symmetry breaking. 
\end{description}

\section{Diagram by diagram}

We will use the following letters to denote algebra elements and
unitaries: Let
$\aaa=M_A(\cc)\op M_B(\cc)\op M_C(\cc)\op M_D(\cc)\owns
(a,b,c,d)$. 
The extended lift is defined by
\bb L(u,v,w,z):=\rho (\hat u,\hat v,\hat w,\hat z)\,J \rho (\hat u,\hat
v,\hat w,\hat z) J^{-1}\ee
with
\bb \hat u&:=&u\,(\det u)^{q_{11}}(\det v)^{q_{12}}(\det
w)^{q_{13}}( \det z)^{q_{14}}\in U(A),\\
\hat v&:=&v\,(\det u)^{q_{21}}(\det v)^{q_{22}}(\det w)^{q_{23}}
( \det z)^{q_{24}}\in
U(B),\\
\hat w&:=&w\,(\det u)^{q_{31}}(\det v)^{q_{32}}(\det w)^{q_{33}}
( \det z)^{q_{34}}\in U(C),\\
 \hat z&:=&z\,(\det u)^{q_{41}}(\det v)^{q_{42}}(\det w)^{q_{43}}
( \det z)^{q_{44}}\in U(D),\ee
and unitaries $(u,v,w,z)\in U( M_A(\cc)\op M_B(\cc)\op M_C(\cc)\op
M_D(\cc) )$. It is understood that for instance if $A=1$ we set
$u=1$ and $q_{j1}=0$, 
$j=1,2,3,4$. If
$M_A(\cc)$ is replaced by $M_A(\rr)$ or $M_{A/2}(\hhh)$
 we set $q_{j1}=0$ and
$q_{1j}=0$. 
\vspace{1\baselineskip}

{\bf Diagram 1} yields:
\bb \rho _L =\pp{b\ot 1_A&0\cr  0&d\ot 1_C},&&
\rho _R=\pp{a\ot 1_A&0\cr  0&^\alpha a\ot 1_C},\cr \cr \cr
\rho _L^c=\pp{1_B\ot\,^{\alpha '} a&0\cr  0&1_D\ot c},&&
\rho _R^c=\pp{1_A\ot \,^{\alpha '}a&0\cr  0&1_A\ot c},\ee
 and
\bb \mm=\pp{M_1\ot 1_A&0\cr  0&M_2\ot1_C},\qq M_1\in
M_{B\times
A}(\cc),\ M_2\in M_{D\times A}(\cc).\ee
The parameters $\alpha $ and $\alpha '$ take values $\pm 1$ and
distinguish between fundamental representation and its complex
conjugate: $^1 a:=a$, $^{-1}a:=\bar a$.
 The colour algebras consist of $a$s and $c$s. The $a$s
are broken and therefore we must have $A=1$. We want at most one
massless Weyl fermion, which leaves us with three possibilities. The
first is
$B=2,\ D=1$.
The fluctuations read:
\bb\varphi _1  =\sum_j r_j\,\hat v _jM_1 \hat u_j^{-1},\qq
\varphi _2  =\sum_j r_j\,\hat z _jM_2\,^\alpha  \hat u_j^{-1}.\ee
We can decouple the two scalars $\varphi _1$ and $\varphi _2$ by
means of the fluctuation:
$ r_1={\textstyle\frac{1}{2}},$ $\hat u_1=1,$ $\hat v_1=1_2,$
$\hat w_1=1_C$, $\hat z_1=1$,
$ r_2={\textstyle\frac{1}{2}},$ $\hat u_2=1,$ $\hat v_2=-1_2,$
$\hat w_2=1_C$, $\hat z_2=1$.  Since the arrows $M_1$ and $M_2$ are
disconnected, the Higgs potential is a sum of a potential in $\varphi
_1$ and of a potential in $\varphi _2$.  
The minimum is
$\stackrel{\circ}{\varphi
}_1=\mu{(4\lambda})^{-\frac{1}{2}}\pp{1\cr 0}$  and
$\stackrel{\circ}{\varphi }_2=\mu{(4\lambda})^{-\frac{1}{2}}$ and
the model is dynamically degenerate. The same accidents happen for
the second possibility, $B=1,\ D=2$. We are left with the third
possibility, $B=D=1$.
Anomaly cancellations imply that the fermion couplings are vectorial:
$q_{13}=q_{23}$, $q_{43}=\alpha q_{13}$, and
there is no spontaneous symmetry breaking,
$SU(C)\times U(1)\,\rightarrow\, SU(C)\times U(1)$. The model
describes electro-strong forces with one charged lepton, one
quark and no scalar. The masses of both fermions are arbitrary but
nonvanishing, the electric charges are arbitrary, nonvanishing for
the lepton, and the number of colours $C$ is arbitrary, $C\ge 2$.
\vspace{1\baselineskip}

 {\bf Diagrams 3, 4} and {\bf 7 } are treated in the same way
and produce only the electro-strong model.
\vspace{1\baselineskip}

{\bf Diagram 2} yields:
\bb \rho _L =\pp{b\ot 1_A&0\cr  0&^\alpha a\ot 1_C},&&
\rho _R=\pp{a\ot 1_A&0\cr  0& d\ot 1_C},\cr \cr \cr
\rho _L^c=\pp{1_B\ot\,^{\alpha '} a&0\cr  0&1_A\ot c},&&
\rho _R^c=\pp{1_A\ot \,^{\alpha '}a&0\cr  0&1_D\ot c},\ee
 and
\bb \mm=\pp{M_1\ot 1_A&0\cr  1_A\ot M_3&M_2\ot1_C},\qq M_1\in
M_{B\times
A}(\cc),\ M_2\in M_{A\times D}(\cc),\ M_3\in M_{C\times A}(\cc).\ee
If $M_3=0$ diagram 2 is treated as diagram 1. We consider the case
$M_3\not= 0$. Then the first order axiom implies $\alpha =1$.
 The colour algebras consist of $a$s and $c$s. Both are broken and
we take $A=C=1$. Then $\mm$ is of rank two or less. If we want at most
one massless Weyl fermion, we must take
$B=2$ and $D=1$ or $B=1$ and $D=2$. Anomaly cancellations then imply
that the doublet of fermions does not couple to the determinant of the
$2\times 2$ unitary, `the hypercharge of the doublet is zero'. On the
other hand the little group turns out either trivial or $U(1)$. In the
latter case the neutrino sitting in the doublet is charged under this
$U(1)$. 
\vspace{1\baselineskip}

{\bf Diagram 5} has no unbroken colour and fails as the first
possibility of diagram 1, $B=2,\ D=1$. 
\vspace{1\baselineskip}

{\bf Diagram 6} has colour $a$ and $b$, both are broken implying
$A=B=1$. Then we must have $C=D=1$ to avoid two or more
neutrinos.
\vspace{1\baselineskip}

{\bf Diagram 8}  yields
\bb \rho _L =\pp{a\ot 1_C&0\cr  0&c\ot 1_B},&&
\rho _R=\pp{a\ot 1_A&0\cr  0& d\ot 1_C},\cr \cr \cr
\rho _L^c=\pp{1_A\ot\,^{\gamma } c&0\cr  0&1_C\ot b},&&
\rho _R^c=\pp{1_A\ot b&0\cr  0&1_D\ot b},\ee
 and
\bb \mm=\pp{1_A\ot M_1&0\cr  M_3\ot 1_B&M_2\ot1_B},\qq M_1\in
M_{C\times
B}(\cc),\ M_2\in M_{C\times D}(\cc),\ M_3\in M_{C\times
A}(\cc).\ee
We suppose that $M_3$ does not vanish otherwise diagram 8 is treated as
diagram 1.  Broken colour implies $A=B=1$. Neutrino counting leaves two
possibilities, $C=1$ and $D=2$ or $C=D=2$. The first possibility is disposed
of as in diagram 2. For the second possibility, anomaly cancellations
imply that the determinants of the $2\times 2$ unitaries $w$ and $z$ do
not couple to the right-handed fermions. 
\vspace{1\baselineskip}

{\bf Diagrams 14, 15, 16} and {\bf 17} share the fate of diagram 6.
\vspace{1\baselineskip}

{\bf Diagram 9} yields
\bb \rho _L =\pp{b\ot 1_A&0&0\cr  0&c\ot 1_B&0\cr 
0&0&d\ot 1_D},&&
\rho _R=\pp{a\ot 1_A&0&0\cr  0&^{\beta _1} b\ot 1_B&0\cr 
0&0&^{\beta _2} b\ot 1_D},\cr
\cr
\cr
\rho _L^c=\pp{1_B\ot\,^{\alpha '} a&0&0\cr  0&1_C\ot \,^{\beta
'}b&0\cr 0&0&1_D\ot\,^{\delta '}d},&&
\rho _R^c=\pp{1_A\ot \,^{\alpha '}a&0&0\cr  0&1_B\ot \,^{\beta
'}b&0\cr 0&0&1_B\ot\,^{\delta '}d},\eee
 and
\bb \mm=\pp{M_1\ot 1_A&1_B\ot M_4&1_B\ot M_5\cr 
0&M_2\ot1_B&0\cr  0&0&M_3\ot 1_D}\ee
with 
$M_1\in M_{B\times
A}(\cc),\ M_2\in M_{C\times D}(\cc),\ M_3\in M_{D\times B}(\cc),
\ M_4\in M_{A\times B}(\cc),\ M_5\in M_{A\times D}(\cc).$ 
If $M_4$ is nonzero $\beta _1$ must
be one and if $M_5$ is nonzero $\beta _2$ must be one. From broken
colour and neutrino counting we have
$A=B=D=1$ and $C=2$.  Vanishing anomalies imply that the first and the
third fermion have vectorlike hypercharges, while the hypercharges
of the second fermion are zero. It is therefore neutral under the little
group $U(1)$.
\vspace{1\baselineskip}

{\bf Diagrams 10, 11, 12} and {\bf 13} fail in the same way.
\vspace{1\baselineskip}

{\bf Diagram 18} produces the following triple:
\bb \rho _L =\pp{a\ot 1_C&0\cr  0&^\alpha a\ot 1_D},&&
\rho _R=\pp{b\ot 1_C&0&0\cr  0&^{\beta _1}\bar b\ot 1_C&0\cr 
0&0&^{\beta _2} \bar b\ot 1_D},\cr
\cr
\cr
\rho _L^c=\pp{1_A\ot c&0\cr  0&1_A\ot d\cr
},&&
\rho _R^c=\pp{1_B\ot c&0&0\cr  0&1_B\ot c&0\cr
0&0&1_B\ot d},\eee
 and
\bb \mm=\pp{M_1\ot 1_C&M_2\ot 1_C&0\cr 
0&0&M_3\ot1_D},\qq
M_1,\ M_2,\ M_3\in M_{A\times
B}(\cc).\ee
Counting neutrinos leaves two possibilities, $A=2$ and 
$B=1$, or $A=1$ and $B=1$. If the neutrino is to be neutral under
the little group we must have $D=1$ for the first possibility and
$C=1$ for the second.

The first possibility has two $U(1)$s if $C\ge 2$ and all four algebras
consist of matrices with complex entries. They are parameterized by
$\det u$ and by $\det w$. For $(\alpha ,\beta _1,\beta _2)=+++,\ 
-++,\ ++-$, and $-+-$,
anomaly cancellations imply that the
fermionic hypercharges of both $U(1)$s are proportional:
\bb
q_{11}=-{\textstyle\frac{1}{2}} ,\ q_{21}=Cp, &&
q_{31}=p,\qq\qq\  q_{41}=-Cp,\\ 
q_{13}=0,\qq\ q_{23}=Cq, &&
q_{33}=q-{\textstyle\frac{1}{C}},\ q_{43}=-Cq,\ee
with $p,q\in\qqq$.
Consequently one linear combination of the two generators decouples
from the fermions and is absent from the spectral action. This is
similar to what happens to the scalar in the electro-strong model of
diagram 1. The hypercharges of the remaining generator are those of
the standard model with $C$ colours. In the remaining four cases
$(\alpha ,\beta _1,\beta _2)=+-+,\ --+,\ +--,$ and$---$, all 10
fermionic hypercharges vanish and the three leptons are neutral
under the little group. Finally the cases where some of the four summands
consist of matrices with real or quaternionic entries are treated as in the
situation with three summands and they produce the same submodels of
the standard model.

The first possibility with $C=1$ has only one $U(1)$. For the four sign
assignments: $(\alpha ,\beta _1,\beta _2)=+++,\ 
-++,\ ++-$, and $-+-$, we find anomaly free lifts with non-trivial little
group. E.g. for the first assignment, we get $q_{11}=-1/2$,
$q_{21}=p$, $q_{31}=-p$, and $q_{41}=-p$, with $p\in\qqq$. All
four assignments produce the electro-weak model
$(SU(2)\times U(1))/\zz_2\rightarrow U(1)$ of protons, neutrons,
neutrinos and electrons.

 The second possibility, $A=B=C=1$, contains at least one 
chiral lepton with vanishing hypercharge and therefore neutral
under the little group. 
 
\vspace{1\baselineskip}

{\bf Diagram 19} 
gives the same results as diagram 18.
\vspace{1\baselineskip}

In {\bf diagrams 20, 21} and {\bf 22}, the elements $a$ and $b$
must be matrices with complex entries to allow for conjugate
representations. Without both the fundamental representation and
its complex conjugate these diagrams would violate the condition
that every nonvanishing entry of the multiplicity matrix of a (blown
up) Krajewski diagram must have the same sign as its transposed
element if the latter is not zero. Diagram 20 is treated and fails as
diagram 2, diagrams 21 and 22 are treated and fail as diagram 13.

At this point we have exhausted all diagrams of figure 2.

\section{Conclusion and outlook}

For three summands there was essentially one irreducible spectral triple
satisfying all items on our shopping list: the standard model with one
generation of quarks and leptons and an arbitrary number of colours
(greater than or equal to two).  We say `essentially' because  a few
submodels of the standard model can be obtained. Going to four summands
adds precisely one model, the electro-strong model for one massive quark
with an arbitrary number of colours and one massive, charged
lepton. Note that this is the first appearance of a spectral triple without
any Higgs scalar and without symmetry breaking.

We still have to prove that spectral triples in four summands involving
letter preserving arrows like $a\ \bar a$ do not produce any
model compatible with our shopping list. 

We are curious to know what happens in five (and more) summands.
Here ladder diagrams do not exist and we may speculate that our present
list of models exhausts our shopping list also in any number of
summands.

Anyhow two questions remain: Who ordered  three colours? Who
ordered three generations?

\vfil\eject
\enlargethispage{1cm}
\thispagestyle{empty}

\begin{tabular}{ccc}
\rxyg{0.7}{
,(5,-5);(10,-5)**\dir{-}?(.6)*\dir{>}
,(5,-15);(20,-15)**\crv{(13,-18)}?(.6)*\dir{>}
}   & $\;\;\;$$\;\;\;$

\rxyg{0.7}{
,(5,-5);(10,-5)**\dir{-}?(.6)*\dir{>}
,(20,-15);(5,-15)**\crv{(13,-18)}?(.6)*\dir{>}
,(5,-5);(5,-15)**\crv{(8,-10)}?(.6)*\dir{>}
} & $\;\;\;$$\;\;\;$

\rxyg{0.7}{
,(5,-5);(10,-5)**\dir{-}?(.6)*\dir{>}
,(10,-15);(20,-15)**\crv{(15,-18)}?(.6)*\dir{>}
}
\\
\\ diag. 1 &  $\;$$\;$ diag. 2 &  $\;$$\;$ diag. 3
\\
\\

\rxyg{0.7}{
,(5,-5);(10,-5)**\dir{-}?(.6)*\dir{>}
,(20,-15);(10,-15)**\crv{(15,-18)}?(.6)*\dir{>}
}   & $\;\;\;$$\;\;\;$

\rxyg{0.7}{
,(10,-5);(15,-5)**\dir{-}?(.6)*\dir{>}
,(5,-10);(20,-10)**\crv{(13,-13)}?(.6)*\dir{>}
} & $\;\;\;$$\;\;\;$

\rxyg{0.7}{
,(10,-5);(15,-5)**\dir{-}?(.6)*\dir{>}
,(10,-5);(10,-20)**\crv{(7,-13)}?(.6)*\dir{>}
}
\\
\\ diag. 4 &  $\;$$\;$ diag. 5 &  $\;$$\;$ diag. 6
\\
\\

\rxyg{0.7}{
,(10,-5);(15,-5)**\dir{-}?(.6)*\dir{>}
,(15,-10);(20,-10)**\dir{-}?(.6)*\dir{>}
} & $\;\;\;$$\;\;\;$

\rxyg{0.7}{
,(5,-10);(5,-15)**\dir{-}?(.6)*\dir{>}
,(5,-10);(15,-10)**\crv{(10,-13)}?(.6)*\dir{>}
,(20,-10);(15,-10)**\dir{-}?(.6)*\dir{>}
} & $\;\;\;$$\;\;\;$

\rxyg{0.7}{
,(5,-5);(10,-5)**\dir{-}?(.6)*\dir{>}
,(10,-10);(15,-10)**\dir{-}?(.6)*\dir{>}
,(10,-20);(20,-20)**\crv{(15,-17)}?(.6)*\dir{>}
,(10,-20);(10,-5)**\crv{(7,-12.5)}?(.6)*\dir{>}
,(10,-10);(10,-5)**\dir{-}?(.6)*\dir{>}
}
\\
\\ diag. 7 &  $\;$$\;$ diag. 8 &  $\;$$\;$ diag. 9
\\
\\
\rxyg{0.7}{
,(5,-5);(10,-5)**\dir{-}?(.6)*\dir{>}
,(10,-10);(10,-5)**\dir{-}?(.6)*\dir{>}
,(10,-10);(15,-10)**\dir{-}?(.6)*\dir{>}
,(20,-20);(10,-20)**\crv{(15,-17)}?(.6)*\dir{>}
,(10,-10);(10,-20)**\crv{(13,-15)}?(.6)*\dir{>}
} & $\;\;\;$$\;\;\;$

\rxyg{0.7}{
,(5,-5);(5,-10)**\dir{-}?(.6)*\dir{>}
,(15,-10);(20,-10)**\dir{-}?(.6)*\dir{>}
,(15,-10);(10,-10)**\dir{-}?(.6)*\dir{>}
,(20,-20);(20,-10)**\crv{(17,-15)}?(.6)*\dir{>}
,(15,-10);(5,-10)**\crv{(10,-12)}?(.6)*\dir{>}
}   & $\;\;\;$$\;\;\;$

\rxyg{0.7}{
,(5,-5);(10,-5)**\dir{-}?(.6)*\dir{>}
,(5,-5);(5,-10)**\dir{-}?(.6)*\dir{>}
,(15,-10);(20,-10)**\dir{-}?(.6)*\dir{>}
,(15,-10);(5,-10)**\crv{(10,-13)}?(.6)*\dir{>}
,(20,-20);(20,-10)**\crv{(17,-15)}?(.6)*\dir{>}
}
\\
\\ diag. 10 &  $\;$$\;$ diag. 11 &  $\;$$\;$ diag. 12
\\
\end{tabular}

\begin{tabular}{ccc}

\rxyg{0.7}{
,(5,-5);(10,-5)**\dir{-}?(.6)*\dir{>}
,(5,-5);(5,-10)**\dir{-}?(.6)*\dir{>}
,(15,-10);(5,-10)**\crv{(10,-13)}?(.6)*\dir{>}
,(10,-20);(20,-20)**\crv{(15,-17)}?(.6)*\dir{>}
,(10,-20);(10,-5)**\crv{(13,-12.5)}?(.5)*\dir{>}
} & $\;\;\;$$\;\;\;$

\rxyg{0.7}{
,(5,-5);(10,-5)**\dir{-}?(.6)*\dir{>}
,(10,-15);(10,-5)**\crv{(12,-10)}?(.5)*\dir{>}
,(10,-20);(20,-20)**\crv{(15,-17)}?(.6)*\dir{>}
,(10,-20);(10,-5)**\crv{(7,-12.5)}?(.6)*\dir{>}
}   & $\;\;\;$$\;\;\;$

\rxyg{0.7}{
,(5,-5);(10,-5)**\dir{-}?(.6)*\dir{>}
,(10,-15);(10,-20)**\dir{-}?(.6)*\dir{>}
,(10,-15);(10,-5)**\crv{(12,-10)}?(.6)*\dir{>}
,(20,-20);(10,-20)**\crv{(15,-17)}?(.6)*\dir{>}
}
\\
\\ diag. 13 &  $\;$$\;$ diag. 14 &  $\;$$\;$ diag. 15
\\
\\

\rxyg{0.7}{
,(5,-5);(10,-5)**\dir{-}?(.6)*\dir{>}
,(10,-15);(15,-15)**\dir{-}?(.6)*\dir{>}
,(10,-20);(20,-20)**\crv{(15,-17)}?(.6)*\dir{>}
} & $\;\;\;$$\;\;\;$

\rxyg{0.7}{
,(5,-5);(10,-5)**\dir{-}?(.6)*\dir{>}
,(15,-15);(10,-15)**\dir{-}?(.6)*\dir{>}
,(20,-20);(10,-20)**\crv{(15,-17)}?(.6)*\dir{>}
} & $\;\;\;$$\;\;\;$

\rxyg{0.7}{
,(5,-15)*\cir(0.4,0){}*\frm{*}
,(10,-15);(5,-15)**\dir2{-}?(.6)*\dir{>}
,(10,-20);(5,-20)**\dir{-}?(.6)*\dir{>}
}
\\
\\ diag. 16 &  $\;$$\;$ diag. 17 &  $\;$$\;$ diag. 18
\\
\\

\rxyg{0.7}{
,(5,-15)*\cir(0.4,0){}*\frm{*}
,(10,-15);(5,-15)**\dir2{-}?(.6)*\dir{>}
,(5,-20);(10,-20)**\dir{-}?(.6)*\dir{>}
} & $\;\;\;$$\;\;\;$

\rxyg{0.7}{
,(10,-5);(15,-5)**\dir{-}?(.6)*\dir{>}
,(20,-10);(5,-10)**\crv{(12.5,-13)}?(.6)*\dir{>}
} & $\;\;\;$$\;\;\;$

\rxyg{0.7}{
,(5,-5);(10,-5)**\dir{-}?(.6)*\dir{>}
,(5,-10);(15,-10)**\crv{(10,-13)}?(.6)*\dir{>}
,(10,-20);(20,-20)**\crv{(15,-17)}?(.6)*\dir{>}
,(10,-20);(10,-5)**\crv{(13,-12.5)}?(.5)*\dir{>}
}
\\
\\ diag. 19 &  $\;$$\;$ diag. 20 &  $\;$$\;$ diag. 21
\\
\\

\rxyg{0.7}{
,(5,-5);(10,-5)**\dir{-}?(.6)*\dir{>}
,(5,-10);(15,-10)**\crv{(10,-13)}?(.6)*\dir{>}
,(20,-20);(20,-10)**\crv{(17,-15)}?(.6)*\dir{>}
,(5,-10);(20,-10)**\crv{(12.5,-7)}?(.6)*\dir{>}
} & $\;\;\;$$\;\;\;$

{}
 & $\;\;\;$$\;\;\;$

{}

\\
\\ diag. 22 &  &
\\

\end{tabular}
\enlargethispage{1cm}

\end{document}